\documentclass[aps, 12pt, showkeys, nofootinbib]{revtex4}

\newcommand{\be}{\begin{eqnarray}}
\newcommand{\ee}{\end{eqnarray}}
\newcommand{\ba}{\begin{array}}
\newcommand{\ea}{\end{array}}

\newcommand{\nn}{\nonumber}
\newcommand{\pa}[1]{\left(#1\right)}
\newcommand{\paq}[1]{\left[#1\right]}
\newcommand{\pag}[1]{\left\{#1\right\}}

\newcommand{\K}{\mathbf{k}}

\newcommand{\Q}{{\mathbf q}}

\newcommand{\X}{{\mathbf x}}

\usepackage[english]{babel}
\usepackage{amsmath,amssymb}
\usepackage{graphicx,color}
\usepackage{braket}
\usepackage{hyperref}
\usepackage[normalem]{ulem}
\usepackage{cancel}
\usepackage{tikz}
\usetikzlibrary{snakes,decorations.pathmorphing}

\begin{document}

\title{Angular momentum tail contributions\\ to compact binary dynamics}

\author{Gabriel Luz Almeida}
\email{galmeida@seu.edu.cn}
\affiliation{School of Physics \& Shing-Tung Yau Center, Southeast University, Nanjing 210096, China}

\author{Alan M\"{u}ller}
\email{alan.muller@unesp.br}
\affiliation{Instituto de F\'\i sica Te\'orica, UNESP - Universidade Estadual Paulista, Sao Paulo 01140-070, SP, Brazil}

\author{Stefano Foffa}
\email{stefano.foffa@unige.ch}
\affiliation{D\'epartement de Physique Th\'eorique and Gravitational Wave Science Center, Universit\'e de Gen\`eve, CH-1211 Geneva, Switzerland}

\author{Riccardo Sturani}
\email{riccardo.sturani@unesp.br}
\affiliation{Instituto de F\'\i sica Te\'orica, UNESP - Universidade Estadual Paulista, Sao Paulo 01140-070, SP, Brazil \\ ICTP South American Institute for Fundamental Research, Sao Paulo 01140-070, SP, Brazil}

\begin{abstract}
We derive the effective action governing the dynamics of a compact binary system when gravitational radiation is emitted by any mass or current multipole,
scattered by the quasi-static field associated with the binary's angular momentum, and then reabsorbed.
Among such angular momentum failed-tail processes, the ones involving multipole moments up to mass and current octupoles, which mix also
with quadrupoles of opposite parity, contribute to the system dynamics
at sixth post-Newtonian order; we display these terms explicitly as a particular case of our general derivation.
Additionally, we derive the radiative
multipole moments associated to arbitrary angular momentum failed-tails in emission processes.
\end{abstract}


\maketitle

\nopagebreak

\section{Introduction}

The two-body problem in General Relativity has been extensively investigated by generations of researchers.
If the post-Newtonian (PN) expansion \cite{Blanchet:2013haa} has been traditionally the privileged analytical method for phenomenological applications \cite{Will:2014kxa}, other approximation schemes, like the post-Minkowskian \cite{Westpfahl:1979gu,Damour:2016gwp,Damour:2019lcq} or the self-force \cite{Poisson:2011nh} approach, 
are now widely used to uncover new sectors of the parameter space and to study different dynamical configurations. Moreover, all those different schemes are employed
for cross-checking and to validate each other in the overlapping domains of validity, as well as for comparisons to numerical relativity results \cite{Rettegno:2023ghr,Boyle:2019kee}.

Still, the PN framework remains
central for most of the phenomenologically relevant events and candidate events for second generation detectors
\cite{TheLIGOScientific:2014jea, TheVirgo:2014hva, KAGRA:2020tym} and future third generation \cite{Maggiore:2019uih,Reitze:2019iox} and space \cite{LISA:2022kgy,TianQin:2020hid}
gravitational wave (GW) detectors as it is at the base of waveforms currently used to filter data \cite{Hamilton:2025xru,Estelles:2025zah}. Even if accuracy requirements
are usually expressed in terms of scalar products of waveforms (or \emph{matches}) rather than in PN orders \cite{Purrer:2019jcp,Thompson:2025hhc},
it is expected that a deeper and more analytic description help is a welcome
ingredient for waveform construction as for hybrid models \cite{Varma:2018mmi}.

Within the PN scheme, even and odd levels in powers of
  $v/c$ are clearly separate up to the fourth perturbative order, the former describing the conservative dynamics due to the exchange of quasi-instantaneous potential modes, the latter the backreaction from radiation emission, which starts at order $\pa{v/c}^5$ (or 2.5PN) in the equations of motion.

The first instance of mixing between potential and radiative effects happens at 4PN (completely determined since a decade \cite{Damour:2014jta,Marchand:2017pir,Foffa:2019yfl}) where, besides the potential modes, also some hereditary effects called tails \cite{Blanchet:1987wq} (given by the backscattering of the emitted radiation with the Newtonian field generated by mass of the binary itself)
contribute to the dynamics.

The current frontier of our knowledge is at 5PN where, besides the (next order of the) already mentioned potential \cite{Blumlein:2020pyo} and tail terms \cite{Foffa:2019eeb,Henry:2023sdy,Almeida:2023yia}, a new hereditary effect, called memory (the backscattering of the emitted radiation by other gravitational radiation) sets in. The impact of the memory on the binary dynamics has been recently uncovered  in \cite{Porto:2024cwd,Almeida:2024lbv,Dlapa:2026oyq}.

With the 5PN level complete, it is natural to look forward and assess the situation at 6PN\footnote{Also the 5.5PN level present an interesting mix, between the known NNNLO  purely radiative effects and a new class of hereditary terms, the tails of tails \cite{Detweiler:2008ft,Edison:2022cdu}.}: the potential sector was known up to $G^4$ \cite{Blumlein:2021txj} until
recently, but in a recent work \cite{Brunello:2025gpf} the static, $G^7$ sector has been completely determined. 
This achievement, made possible by overcoming the technical challenges involving the integration by parts for 6-loop Feynman integrals as well as the computation of new master integrals, changed the perspective on the feasibility of the entire 6PN sector, which can now be considered within reach.

This work gives another impulse in this direction by determining the whole tower of angular momentum tails, that is the analog of mass tails (the one appearing first at 4PN), where the Newtonian field which backscatters the radiation is generated by the total angular momentum instead of
the mass of the binary. The angular momentum tail manifests itself
already at 5PN \cite{Henry:2023sdy,Almeida:2023yia,Porto:2024cwd}, but it is at
6PN that its distinctive feature, i.e. the interference between multipoles of different order and parity, makes its first
appearance.

The plan of the paper is straightforward. In section \ref{se:setup} we cover some methodological background and we give a simple explanation of the interference; in doing so we provide a general expression for the contribution of angular momentum tails to the radiative moments associated to the GW at infinity. In section \ref{se:results} we show and discuss our results; our computation is not actually limited at 6PN, but it is valid in general for every possible radiative multipole involved, and we extract explicitly the 6PN subsector as a particular case. Our conclusions are contained in section \ref{se:conc}.

\section{Setup}\label{se:setup}
We describe the binary system coupled to gravity via the usual Einstein-Hilbert action (fixed to the de Donder gauge) plus the tower of linear multipolar couplings\footnote{Here $\Lambda\equiv \pa{32 \pi G_d}^{-1/2}$, $G_d$ being Newton's constant in $d+1$-dimensions;
$\Gamma^\mu\equiv\Gamma^\mu_{\nu\rho}g^{\nu\rho}$, with $\Gamma^\mu_{\nu\rho}$ the
standard Christoffel connection. We adopt the notation $\int_t\equiv \int{\rm d}t$, and later below $\int_\K\equiv \int \frac{{\rm d}^{d}k}{\pa{2\pi}^d}$, and our metric convention is ``mostly plus'':
  $\eta_{\mu\nu}={\rm diag}\pa{-1,+1,+1,+1}$. The collective index $L-2$ stands for $ i_1\dots i_{{\ell}-2}$ while the $J^{k|iL-2l}$ are the $d$-dimensional generalizations \cite{Henry:2021cek} of the 3-dimensional magnetic(current)-type multipoles
$J^{ijL-2}=\frac12 \epsilon_{kl(i} \left.J^{k|jL-2)l}\right|_{d=3}$, $L^i =\frac12 \epsilon_{ikl}\left.J^{k|l}\right|_{d=3}$. }:
\be\label{eq:sgrav}
{\cal S}={\cal S}_{\rm EH+GF}+{\cal S}_{\rm mult}\,,\quad{\rm with}\quad {\cal S}_{\rm EH+GF}=2\Lambda^2 \int {\rm d}^{d+1}x \sqrt{-g} \left[ {\cal R}(g) - \frac12 \Gamma_\mu \Gamma^\mu \right]\,,
\ee
and
\be
\label{eq:smult}
{\cal S}_{\rm mult} &=& \int_t\,\left[ \frac12 E h_{00} -\frac12 J^{k|l} h_{0k,l} - \sum_{\ell \ge 2}\left( c_\ell^{(I)} I^{ijL-2} \partial_{L-2} {\cal R}_{0i0j} + \frac{c_\ell^{(J)}}{2} J^{k|iL-2l} \partial_{L-2} {\cal R}_{0ilk} \right) \right]\,,\\
&&{\rm with}\qquad c_\ell^{(I)} = \frac{1}{\ell!}\,, \qquad c_\ell^{(J)} = \frac{2\ell}{(\ell+1)!}\,,\nn
\ee
where eq.(\ref{eq:smult}) is considered only at linear level in the gravitational perturbation $h_{\mu\nu}$.

\begin{figure}[h]
    \begin{tikzpicture}
      	\draw [black, thick] (0,0) -- (6,0);
      	\draw [black, thick] (0,-0.1) -- (6,-0.1);
      	\draw[decorate, decoration=snake, line width=1.5pt, green] (0.5,0) arc (180:0:2.5);
      	\draw[red, thick, dotted] (3.,0) --  (3.,2.5) ;
      	\filldraw[black] (0.5,-0.1) circle (3pt) node[anchor=north] {$I_\ell$};
      	\filldraw[black] (5.5,-0.1) circle (3pt) node[anchor=north] {$I_\ell$};
	\filldraw[black] (3,-0.1) circle (3pt) node[anchor=north] {${\bf L}$};
    \end{tikzpicture}
        \begin{tikzpicture}
      	\draw [black, thick] (0,0) -- (6,0);
      	\draw [black, thick] (0,-0.1) -- (6,-0.1);
      	\draw[decorate, decoration=snake, line width=1.5pt, green] (0.5,0) arc (180:0:2.5);
      	\draw[red, thick, dotted] (3.,0) --  (3.,2.5) ;
      	\filldraw[black] (0.5,-0.1) circle (3pt) node[anchor=north] {$J_\ell$};
      	\filldraw[black] (5.5,-0.1) circle (3pt) node[anchor=north] {$J_\ell$};
	\filldraw[black] (3,-0.1) circle (3pt) node[anchor=north] {${\bf L}$};
    \end{tikzpicture}\\
        \begin{tikzpicture}
      	\draw [black, thick] (0,0) -- (6,0);
      	\draw [black, thick] (0,-0.1) -- (6,-0.1);
      	\draw[decorate, decoration=snake, line width=1.5pt, green] (0.5,0) arc (180:0:2.5);
      	\draw[red, thick, dotted] (3.,0) --  (3.,2.5) ;
      	\filldraw[black] (0.5,-0.1) circle (3pt) node[anchor=north] {$I_\ell$};
      	\filldraw[black] (5.5,-0.1) circle (3pt) node[anchor=north] {$J_{\ell+1}$};
	\filldraw[black] (3,-0.1) circle (3pt) node[anchor=north] {${\bf L}$};
    \end{tikzpicture}
        \begin{tikzpicture}
      	\draw [black, thick] (0,0) -- (6,0);
      	\draw [black, thick] (0,-0.1) -- (6,-0.1);
      	\draw[decorate, decoration=snake, line width=1.5pt, green] (0.5,0) arc (180:0:2.5);
      	\draw[red, thick, dotted] (3.,0) --  (3.,2.5) ;
      	\filldraw[black] (0.5,-0.1) circle (3pt) node[anchor=north] {$J_\ell$};
      	\filldraw[black] (5.5,-0.1) circle (3pt) node[anchor=north] {$I_{\ell+1}$};
	\filldraw[black] (3,-0.1) circle (3pt) node[anchor=north] {${\bf L}$};
    \end{tikzpicture}\\
    \caption{Angular momentum tails.}\label{fig:Lself}    
\end{figure}
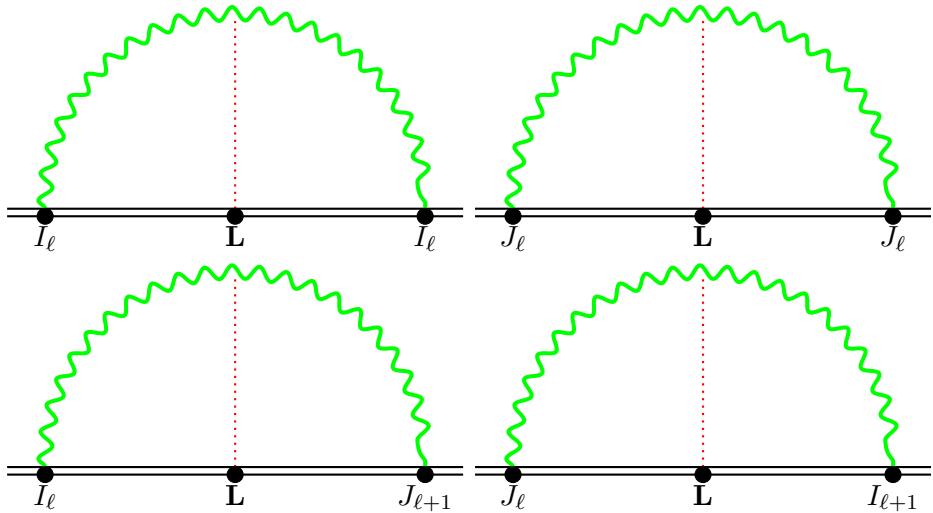

These are the building blocks we are using to compute, via the well consolidated NRGR (Non Relativistic General Relativity) approach \cite{Goldberger:2004jt}, the {\em angular momentum failed-tail} self-interaction diagrams represented in figure \ref{fig:Lself}, which describe the binary dynamics when gravitational radiation is backscattered by the quasi-Newtonian gravitational field associated to its total angular momentum, and then reabsorbed by the binary itself.
The calculation is done in full generality, for any kind (mass or current) and multipolar order associated to the GW radiation, and the process is computed in the same way as the {\em mass tail} diagrams \cite{Almeida:2021xwn}.

Having this said, the result for the angular momentum tail process is different from the mass tail in several aspects. To begin with, it is local (hence the label ``failed") and finite in dimensional regularization: the latter property allows us to write the results directly in terms of the 3-dimensional magnetic multipoles (and the Levi-Civita tensor, when needed)\footnote{Using the
$\epsilon_{ijk}$, as well as the $d=3$ form $J^{ijL-2}$ in expressions that are singular in dim. reg. can result in ambiguity or inconsistencies, see for instance \cite{Almeida:2021jyt}.}.
Moreover, only for the specific case involving two mass quadrupoles, the additional contact diagram shown in figure \ref{fig:contact} contributes to the process.
\begin{figure}[h]
    \begin{tikzpicture}
      	\draw [black, thick] (0,0) -- (6,0);
      	\draw [black, thick] (0,-0.1) -- (6,-0.1);
      	\draw[decorate, decoration=snake, line width=1.5pt, green] (0.5,0) arc (180:0:1.25);
	\draw[decorate, decoration=snake, line width=1.5pt, green] (3,0) arc (180:0:1.25);
      	\filldraw[black] (0.5,-0.1) circle (3pt) node[anchor=north] {$I_{ij}$};
      	\filldraw[black] (5.5,-0.1) circle (3pt) node[anchor=north] {$I_{ij}$};
	\filldraw[black] (3,-0.1) circle (3pt) node[anchor=north] {${\bf L}$};
    \end{tikzpicture}
\caption{Contact diagram.}\label{fig:contact}    
\end{figure}

The computation of such diagram is straightforward due to the factorization
  of the loop integrations, but involves a quadratic coupling which is not included in equation (\ref{eq:smult}); such coupling has been derived in \cite{Porto:2005ac} and can also be recovered from the more general treatment described in \cite{Porto:2024cwd}. The result, which is vanishing for any other multipole different from $I_{ij}$, has been included in equation \ref{eq:Ltail_sem_el} below, for the specific case $\ell=2$.
\footnote{The reason why the contact diagram is vanishing for all the other multipoles is that the quadratic coupling to $J^{k|l}$ can contract only one pair of indices from different multipoles, thus automatically killing all traceless multipoles with $\ell\geq3$.
  The remaining magnetic quadrupole case is vanishing because of parity (it involves a one loop integral with one vector index).}

A final difference between mass and angular momentum tail is that the vectorial nature of $L^i$ generates new contributions to the self-energy sector.
This can be seen by computing the Transverse Traceless (TT) part of the field associated to this process,
which is related to the emission amplitudes $\mathcal{A}^{({\rm e}-L)}\,,\mathcal{A}^{({\rm m}-L)}$, already computed in \cite{Almeida:2023yia}, via the general relation\be
\label{eq:hexpect}
\braket{h_{\mu\nu}(x)}&=& \int\mathcal{D}h\, e^{iS[h]}h_{\mu\nu}(x)=\int_\K \frac{{\rm d}\omega}{2\pi} \frac{e^{-i\omega t+i\K\cdot\X}}{\K^2-(\omega+i0^+)^2}
\frac{{{\cal P}_{\mu\nu}}^{\alpha\beta}}{\Lambda^2}{\cal A}_{\alpha\beta}(\omega,\K)\,,\\
P[h_{\mu\nu},h^{\alpha\beta}]&=&-\frac{i}{\K^2-\omega^2}\frac{{{\cal P}_{\mu\nu}}^{\alpha\beta}}{\Lambda^2}\,,\quad {{\cal P}_{\mu\nu}}^{\alpha\beta}\equiv \frac 12 \pa{\delta_\mu^\alpha\delta_\nu^\beta+\delta_\mu^\beta\delta_\nu^\alpha-\frac 2{d-1}\eta_{\mu\nu}\eta^{\alpha\beta}}.
\ee
The result\footnote{Also for the emission process, the case involving one mass quadrupole gets contribution also from one contact diagram, see \cite{Almeida:2023yia} for details. The expression presented here includes the contribution of such contact diagram for that specific case.}
\be
\label{eq:e-l}
\paq{h^{({\rm e}-L)}_{ij}}^{\text{TT}}  &=&\frac{8r^{-1}}{\left(\ell-1\right)\ell\left(\ell+2\right)!}\times\\
 && \times\paq{2\pa{\ell^2+2}n_{aL-1}\varepsilon_{ab(i}I_{j)L-1}^{\left(\ell+2\right)}L_{b}-(\ell-2)\pa{\ell^2+\ell+4}n_{aL-3}\varepsilon_{abc}L_{b}I_{ijcL-3}^{\left(\ell+2\right)}}^{\text{TT}}\,,\nn\\
\label{eq:e-m}
\paq{h^{({\rm m}-L)}_{ij}}^{\text{TT}}   &=& \frac{16 r^{-1}}{\left(\ell-1\right)\left(\ell+1\right)\left(\ell+3\right)!}\times\left\{\pa{\ell^4+4\ell^3+5\ell^2+2\ell+12} n_{L-1}L_{(i} J_{j)L-1}^{\pa{\ell+2}}\right.\\
&&+\left.\pa{\ell^4+2\ell^3-\ell^2-2\ell+24}\paq{(1-\delta_{\ell2})n_{L-3}J_{ijaL-3}^{\pa{\ell+2}}-n_{aL-2}J_{ijL-2}^{\pa{\ell+2}}}L_a \right\}^{\text{TT}}\nn
\ee
(where $n_{ab}\equiv n_a n_b$, $n_a\equiv-\hat{r}_a$ ),
can then be projected  into radiative moments according to the standard decomposition
\begin{equation}
h_{ij}^{\text{TT}}=4r^{-1}\sum_{\ell\geq2}\frac{1}{\ell!}\left(n_{L-2}{\cal U}_{ijL-2}-\frac{2\ell}{\ell+1}n_{aL-2}\varepsilon_{ab(i}{\cal V}_{j)bL-2}\right)^{\text{TT}}\,.
\label{eq:w}
\end{equation}
This brings to (for details, see Appendix~\ref{app:stf})
\be
{\cal U}_{ijL-2}&\supset& \pa{1-\delta_{\ell2}}U_+(\ell-1)\ L_{\langle i}J_{jL-2\rangle}^{\left(\ell+1\right)}+U_0(\ell)\ L_a \varepsilon_{a b\langle i} I_{j L-2\rangle b}^{\left(\ell+2\right)}
+U_-(\ell+1)\ L_{a} J_{aijL-2}^{\left(\ell+3\right)} \,, \nn\\
{\cal V}_{ijL-2}&\supset&  \pa{1-\delta_{\ell2}} V_+(\ell-1)\ L_{\langle i}I_{jL-2\rangle}^{\left(\ell+1\right)}+V_0(\ell)\ L_a \varepsilon_{ab \langle i} J_{j L-2\rangle b}^{\left(\ell+2\right)}
+V_-(\ell+1)\ L_a I_{aijL-2}^{\left(\ell+3\right)}\,,\nn
\ee
with 
\be\label{eq:radcoeff}
U_-(\ell)&\equiv&  \pa{1-\delta_{\ell2}} \frac{8(\ell+2)\pa{\ell^2 -2 \ell +3}}{(\ell-1)\ell^2(\ell+1)^2(2\ell+1)}\,,\quad V_-(\ell) \equiv-\frac{2(\ell-2)\pa{\ell^2 + 2 \ell +3}}{(\ell-1)^2\ell^2(\ell+1)(2\ell+1)}\,,\nn\\
U_0(\ell)&\equiv&-\frac{2\pa{\ell^4+2 \ell^3-\ell^2-2\ell -12}}{(\ell-1)\ell(\ell+1)^2(\ell+2)}\,,\quad V_0(\ell)\equiv -\frac{2\pa{\ell^4+2 \ell^3-\ell^2-2\ell +12}}{(\ell-1)\ell(\ell+1)^2(\ell+2)}\,,\\
U_+(\ell)&\equiv&\frac{8\pa{\ell^2+4 \ell +6}}{(\ell+1)(\ell+2)(\ell+3)}\,,\quad V_+(\ell) \equiv-\frac{2\pa{\ell^2+2}}{(\ell-1)\ell(\ell+1)}\left[=-\frac{U_+(\ell-2)}4\right]\,.\nn
\ee
This means that, for instance, the angular momentum tail involving a source mass quadrupole $I_{ij}$ generates, besides the expected radiative mass quadrupole ${\cal U}_{ij}$, also a radiative current octupole,  ${\cal V}_{ijk}$ and so on for the other processes.
The above coefficients reproduce and generalize for all $\ell$'s some results already known in the literature, up to $\ell=4$, see \cite{Faye:2014fra}.

This leads to the richer structure in the self energy diagrams reproduced in figure \ref{fig:Lself}, where one can see, in addition to the processes involving two copies of the same moments $I_\ell I_\ell$ and $J_\ell J_\ell$ (like in the mass tails), the appearance of some new contribution coming from mixed terms such as $I_\ell J_{\ell+1}$, $I_{\ell+1} J_\ell$.

\section{Results}\label{se:results}
\begin{figure}[h]
    \begin{tikzpicture}
      	\draw [black, thick] (0,0) -- (4,0);
      	\draw [black, thick] (0,-0.1) -- (4,-0.1);
      	\draw[decorate, decoration=snake, line width=1.5pt, green] (0.5,0) arc (180:0:1.5);
      	\draw[thick, dotted] (2.,0) --  (2.,1.5) ;
      	\filldraw[black] (0.5,-0.1) circle (3pt) node[anchor=north] {$\pa{I_\ell,J_\ell}_+$};
      	\filldraw[black] (3.5,-0.1) circle (3pt) node[anchor=north] {$\pa{I_{\ell'},J_{\ell'}}_-$};
	\filldraw[black] (2,-0.1) circle (3pt) node[anchor=north] {${\bf L}_+$};
	\draw[-stealth,thick] (1., .7)  arc (180:90:.5);
	\draw[-stealth,thick] (2.5, 1.2)  arc (90:0:.5);
    \end{tikzpicture}
    \begin{tikzpicture}
      	\draw [black, thick] (0,0) -- (4,0);
      	\draw [black, thick] (0,-0.1) -- (4,-0.1);
      	\draw[decorate, decoration=snake, line width=1.5pt, green] (0.5,0) arc (180:0:1.5);
      	\draw[thick, dotted] (2.,0) --  (2.,1.5) ;
      	\filldraw[black] (0.5,-0.1) circle (3pt) node[anchor=north] {$\pa{I_\ell,J_\ell}_-$};
      	\filldraw[black] (3.5,-0.1) circle (3pt) node[anchor=north] {$\pa{I_{\ell'},J_{\ell'}}_+$};
	\filldraw[black] (2,-0.1) circle (3pt) node[anchor=north] {${\bf L}_+$};
	\draw[-stealth,thick] (1.5, 1.2)  arc (90:180:.5);
	\draw[-stealth,thick] (3, .6)  arc (0:90:.5);
    \end{tikzpicture}
    \begin{tikzpicture}
      	\draw [black, thick] (0,0) -- (4,0);
      	\draw [black, thick] (0,-0.1) -- (4,-0.1);
      	\draw[decorate, decoration=snake, line width=1.5pt, green] (0.5,0) arc (180:0:1.5);
      	\draw[thick, dotted] (2.,0) --  (2.,1.5) ;
      	\filldraw[black] (0.5,-0.1) circle (3pt) node[anchor=north] {$\pa{I_\ell,J_\ell}_+$};
      	\filldraw[black] (3.5,-0.1) circle (3pt) node[anchor=north] {$\pa{I_{\ell'},J_{\ell'}}_+$};
	\filldraw[black] (2,-0.1) circle (3pt) node[anchor=north] {${\bf L}_-$};
	\draw[-stealth,thick] (1., .7)  arc (180:90:.5);
	\draw[-stealth,thick] (3, .6)  arc (0:90:.5);
    \end{tikzpicture}
    \caption{Keldysh structure of the tail diagrams.}\label{fig:Keldysh}    
\end{figure}
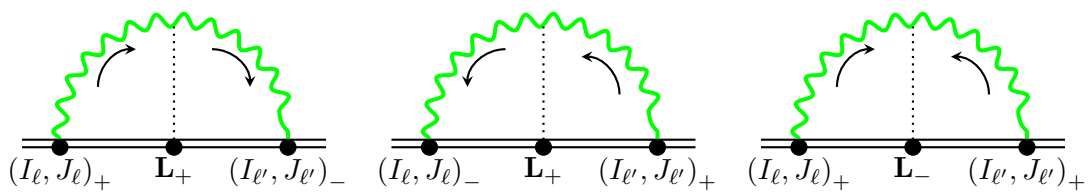

Since the tails involve also dissipative effects, the calculation is done in the in-in formalism using Keldysh variables \cite{Galley:2009px,Keldysh:1964ud}.
Such formalism is necessary for introducing retarded and advanced
  Green functions, whose time orientation is represented in fig.~\ref{fig:Keldysh} by
  arrows. As a result, each term in the original in-out Lagrangean is replaced in the in-in formalism by all possible terms involving the same
  fields and one ``$-$'' label, with the remaining fields being of the ``+''
  type.
  The Green function obtained by the ``contraction'' of a pair of fields $\pag{+-} (\pag{-+})$ is a retarded (advanced) one, and the result of the
  loop integral is sensitive to the Green function boundary condition, as explained in app.~\ref{app:ints}.

The possible configurations are displayed in figure \ref{fig:Keldysh} and they all have the same structure: taking for simplicity one specific process, one has
\be
{\cal S}_{I^\ell {\bf L} J^{\ell+1}}&\equiv&{\cal S}_{I^\ell_+ {\bf L}_+ J^{\ell+1}_-}+{\cal S}_{I^\ell_- {\bf L}_+ J^{\ell+1}_+}+{\cal S}_{I^\ell_+ {\bf L}_- J^{\ell+1}_+}
\nn\\
&=&S^{I^\ell {\bf L} J^{\ell+1}}\paq{I^\ell_+ {\bf L}_+ J^{\ell+1}_- + I^\ell_- {\bf L}_+ J^{\ell+1}_+ - I^\ell_+ {\bf L}_- J^{\ell+1}_+}\,,
\ee
where the (unique) index contraction is implicitly understood.

The effective action is made of a common factor, times a fixed combination of the Keldysh variables, with the relative minus sign due to the different orientation of the retarded propagators.
Thus the determination of the angular momentum tail effect on the binary dynamics reduces to the computation of the four families of coefficients $S^{I^\ell {\bf L} I^\ell}$,
$S^{J^\ell {\bf L} J^\ell}$, $S^{I^\ell {\bf L} J^{\ell+1}}$, $S^{I^{\ell+1} {\bf L} J^\ell}$, for every $\ell$.
The first two families have been already computed in \cite{Almeida:2023yia} and give\footnote{The specific process involving two mass quadrupoles, that is the first equation with $\ell=2$ gets contribution also from a contact diagram, see \cite{Almeida:2023yia} for details. The expression presented here includes the contribution of this contact diagram for that specific case.}
\be
\label{eq:Ltail_sem_el}
{\cal S}_{I^\ell {\bf L} I^\ell}&=&-\frac{G^2\pa{\ell^4+2 \ell^3-\ell^2-2\ell -12}}{(\ell-1)^2\ell^2(\ell+1)!(2\ell+1)!!} \epsilon_{ikl}\int_t \paq{2L^l_+ I_{iL-1,+}^{(\ell+2)}I_{kL-1,-}^{(\ell+1)}-L^l_- I_{iL-1,+}^{(\ell+2)}I_{kL-1,+}^{(\ell+1)}}\,,\\
\label{eq:Ltail_sem_mag}
{\cal S}_{J^\ell {\bf L} J^\ell}&=&-\frac{4 G^2\pa{\ell^4+2 \ell^3-\ell^2-2\ell +12}}{(\ell-1)^2 (\ell+1)^2(\ell+1)!(2\ell+1)!!}\epsilon_{ikl} \int_t \paq{2L^l_+ J_{iL-1,+}^{(\ell+2)}J_{kL-1,-}^{(\ell+1)}-L^l_- J_{iL-1,+}^{(\ell+2)}J_{kL-1,+}^{(\ell+1)}}\,.\nn\\
\ee 
The mixed contributions are computed here for the first time and are
\be
\label{eq:Ltail_smix}
{\cal S}_{I^{\ell+1} {\bf L} J^\ell}&=&\frac{8 G^2(\ell^2+4 \ell+6)}{\ell (\ell+1)^2(\ell+1)!(2\ell+3)!!}\int_t
\paq{L^i_+ I_{iL,-}^{(\ell+2)} J_{L,+}^{(\ell+2)}+L^i_+ I_{iL,+}^{(\ell+2)} J_{L,-}^{(\ell+2)}-L^i_- I_{iL,+}^{(\ell+2)} J_{L,+}^{(\ell+2)}}\,,\nn\\
&&\\
{\cal S}_{I^\ell {\bf L} J^{\ell+1}}&=&-\frac{8 G^2(\ell+3)(\ell^2+2)}{(\ell-1) \ell^2(\ell+2)!(2\ell+3)!!}\int_t 
\paq{L^i_+ I_{L,-}^{(\ell+2)} J_{iL,+}^{(\ell+2)}+L^i_+ I_{L,+}^{(\ell+2)} J_{iL,-}^{(\ell+2)}-L^i_- I_{L,+}^{(\ell+2)} J_{iL,+}^{(\ell+2)}}\,.\nn\\
\ee 
The reader is referred to Appendices ~\ref{app:ints} and \ref{app:unint} for the computational details.

The coefficients of the four families above remind closely the projectors on radiative moments shown in equation (\ref{eq:radcoeff}), and this is a consequence of unitarity.
Indeed one can write, taking the same specific example as above
\be
{\cal S}_{I^\ell_+ {\bf L}_+ J^{\ell+1}_-}&=& \frac {i}{\Lambda^2} \int _{\K} \frac{{\rm d}\omega}{2\pi}\paq{\mathcal{A}^{({\rm e}-L)}_{ij}(\omega,\K)}^{\rm TT}P[h_{ij},h_{kl}]\paq{\mathcal{A}^{\rm (LO,m)}_{kl}(-\omega,-\K)}^{\rm TT}\,,
\ee
with the leading order magnetic amplitude taken from
\be
\label{eq:emLO}
i \paq{\mathcal{A}^{\rm LO}_{ij}(\omega,\K)}^{\rm TT}&=&
\sum_{\ell\geq 2}\frac{(-i)^{\ell-1}}{2} k_{L-2} \left[{c_{\ell}}^{(I)} \omega^2 I^{ijL-2}(\omega)+c_{\ell}^{(J)}\omega k_l \epsilon_{kl(i}  J^{j)kL-2}(\omega)\right]^{\rm TT}
\ee
and the electric tail amplitude $\mathcal{A}^{(e-L)}$ as already computed in \cite{Almeida:2023yia}.

This procedure corresponds to an unitarity cut \cite{Bern:2011qt}
along the right green propagator in the first diagram in figure \ref{fig:Keldysh}; of course it is possible to obtain the same result using the other cut, on the left green propagator, which would result in the glueing of $\mathcal{A}^{\rm (LO,e)}(\omega,\K)$ with $\mathcal{A}^{({\rm m}-L)}(-\omega,-\K)$.
The same logic holds also for all the other processes; indeed, we have followed this strategy as a mean to cross-check the results against the ones obtained above via direct calculation.

From the above results we can recover the only term contributing to 5PN,
\be
{\cal S}_{\rm 5PN}&=&-\frac{G^2}{30}\epsilon_{abc}\int_t \paq{2 L^c_+ I_{aj,+}^{(4)}I_{bj,-}^{(3)}-L^c_- I_{aj,+}^{(4)}I_{bj,+}^{(3)}}\,,
\ee
which has already been computed in \cite{Henry:2023sdy,Almeida:2023yia} (for the conservative part only) and in \cite{Porto:2024cwd}.
We can also explicitly extract the relevant action at 6PN
\be
{\cal S}_{\rm 6PN}&=&-\frac{G^2}{840}\epsilon_{abc}\int_t \paq{2 L^c_+ I_{ajk,+}^{(5)}I_{bjk,-}^{(4)}-L^c_- I_{ajk,+}^{(5)}I_{bjk,+}^{(4)}}\nn\\
&&-\frac{8G^2}{45}\epsilon_{abc}\int_t \paq{2 L^c_+ J_{aj,+}^{(4)}J_{bj,-}^{(3)}-L^c_- J_{aj,+}^{(4)}J_{bj,+}^{(3)}}\nn\\
&&-\frac{G^2}{42}\int_t \paq{L^c_+ I_{ij,+}^{(4)} J_{cij,-}^{(4)}+L^c_+ I_{ij,-}^{(4)} J_{cij,+}^{(4)}-L^c_- I_{ij,+}^{(4)} J_{cij,+}^{(4)}}\nn\\
&&+\frac{4G^2}{315}\int_t \paq{L^c_+ J_{ij,+}^{(4)} I_{cij,-}^{(4)}+L^c_+ J_{ij,-}^{(4)} I_{cij,+}^{(4)}-L^c_- J_{ij,+}^{(4)} I_{cij,+}^{(4)}}\,,
\ee
which should be added to the already known part of the potential \cite{Blumlein:2020znm, Brunello:2025gpf}, and to mass-tail \cite{Almeida:2021xwn} contributions.

For future reference, we display also the 7PN terms:
\be
{\cal S}_{\rm 7PN}&=&-\frac{29G^2}{1360800}\epsilon_{abc}\int_t \paq{2 L^c_+ I_{ajkl,+}^{(6)}I_{bjkl,-}^{(5)}-L^c_- I_{ajkl,+}^{(6)}I_{bjkl,+}^{(5)}}\nn\\
&&-\frac{11G^2}{3360}\epsilon_{abc}\int_t \paq{2 L^c_+ J_{ajk,+}^{(5)}J_{bjk,-}^{(4)}-L^c_- J_{ajk,+}^{(5)}J_{bjk,+}^{(4)}}\nn\\
&&-\frac{11G^2}{42525}\int_t \paq{L^c_+ I_{ijk,+}^{(5)} J_{cijk,-}^{(5)}+L^c_+ I_{ijk,-}^{(5)} J_{cijk,+}^{(5)}-L^c_- I_{ijk,+}^{(5)} J_{cijk,+}^{(5)}}\nn\\
&&+\frac{G^2}{5040}\int_t \paq{L^c_+ J_{ijk,+}^{(5)} I_{cijk,-}^{(5)}+L^c_+ J_{ijk,-}^{(5)} I_{cijk,+}^{(5)}-L^c_- J_{ijk,+}^{(5)} I_{cijk,+}^{(5)}}\,.
\ee

\section{Conclusions}\label{se:conc}
The recent determination of the static part of the 6PN potential removed one of the major obstacles from the completion of this post-Newtonian level.
Here we provide another missing part, i.e. the effective action associated to several angular momentum tails appearing at this perturbative level. The calculation is carried on in full generality and confirms the extended reach of
the NRGR framework to deal with the compact binary dynamics problem.
The 6PN could be completed in a not far future by the determination of (i) the remaining part of the potential (which involves integrals easier then the static part), and (ii) the next-to-leading order memory terms.

\section*{Acknowledgments} 
S.F. is supported by the SwissMap National Center for Competence in Research, and
also acknowledges partial financial support and kind hospitality from the post-graduate program of the Institute of Theoretical Physics (IFT-UNESP) during the
preparation of this work. A.M. acknowledges support by
FAPESP, under grant2023/15479-1 and 2024/10380-0, and acknowledges the hospitality of
Geneva University during the preparation of this work. RS wishes to acknowledge FAPESP
grants n. 2022/06350-2 and 2021/14335-0, as well as CNPq grant n.310165/2021-0. This
work was supported partially by the Start-up Research Fund of Southeast University No.
RF1028624160.

\appendix

\section{STF decomposition}
\label{app:stf}

To derive the contributions from the waveforms (\ref{eq:e-l}) and (\ref{eq:e-m}) to the radiative multipole moments $\mathcal{U}_L$ and $\mathcal{V}_L$ appearing in the waveform decomposition (\ref{eq:w}), we must bring each waveform into the form of a multipole expansion. As explicited in (\ref{eq:w}), this expansion is expressed in terms of symmetric trace-free (STF) tensors (namely $\mathcal{U}_L$ and $\mathcal{V}_L$), i.e., irreducible representations of SO(3) \cite{Ross:2012fc,Thorne:1980ru,Blanchet:2013haa}. However, tensor structures such as $I_{jL-1}^{(\ell+2)}L_b$ and $\varepsilon_{acb}L_b I^{(\ell+2)}_{ijcL-3}$ appearing in (\ref{eq:e-l}) are neither symmetric nor trace-free. To match the multipolar structure of (\ref{eq:w}) and identify the corresponding radiative moments, these tensor structures must be decomposed into STF pieces. We refer to this procedure as \emph{STF decomposition} and explain it in this appendix.

\subsection{Mass-type angular momentum tail}
\label{app:e-stf}

We start from (\ref{eq:e-l}), which we rewrite here in the form
\be\label{eq:e-stf}
\paq{h^{(e-L)}_{ij}}^{\text{TT}} = \frac{r^{-1}}{2}\left( A^e_{\ell} n_{aL-1}\varepsilon_{ab i}I_{j L-1}^{\left(\ell+2\right)}L_{b}+ B_{\ell}^{e}n_{L-2}L_{b}I_{aijL-3}^{\left(\ell+2\right)}\varepsilon_{i_{\ell-2}ab} + i\leftrightarrow j\right)^{\text{TT}}.
\ee
The first term on the rhs of (\ref{eq:e-stf}) resembles the second term of (\ref{eq:w}) and should lead to a contribution $\mathcal{V}_{jbL-1}\propto I_{\langle jL-1}^{(\ell+2)}L_{b\rangle}$. The first thing we must do in order to bring it into the form of a multipole moment is symmetrize over the indices $jbL-1$. To this end, we write the trivial identity
\be\label{eq:identity}
\left(n_{aL-1}\varepsilon_{abi} I_{jL-1}^{(\ell+2)}L_{b}\right)^{\mathrm{TT}}=\left[n_{aL-1}\varepsilon_{abi} \left(I_{(jL-1}^{(\ell+2)}L_{b)}+I_{jL-1}^{(\ell+2)}L_{b}-I_{(jL-1}^{(\ell+2)}L_{b)}\right)\right]^{\mathrm{TT}},
\ee
where we have simply added and subtracted $I_{(jL-1}^{(\ell+2)}L_{b)}$. Then, we note that
\be\label{eq:identity2}
I_{(jL-1}^{(\ell+2)}L_{b)}=\frac{1}{\ell+1}\left(I_{jL-1}^{(\ell+2)}L_{b}+I_{bL-1}^{(\ell+2)}L_{j}+I_{bji_{2}\cdots i_{\ell-2}i_{\ell-1}}^{(\ell+2)}L_{i_{1}}+\cdots+I_{bi_{1}i_{2}\cdots i_{\ell-2}j}^{(\ell+2)}L_{i_{\ell-1}}\right).
\ee
Plugging (\ref{eq:identity2}) into the last term of (\ref{eq:identity}) and noting that the contraction with $n_{aL-1}$ enforces symmetrization over the indices $L-1\equiv i_1\cdots i_{\ell-1}$, one is able to work out the following expression,
\begin{align}\label{eq:identity3}
\begin{split}
\left(n_{aL-1} \varepsilon_{abi} I_{jL-1}^{(\ell+2)}L_{b}\right)^{\mathrm{TT}} & =\biggl(n_{aL-1} \varepsilon_{abi} I_{(jL-1}^{(\ell+2)}L_{b)}\\
&+ \frac{2}{\ell+1} n_{aL-1} \varepsilon_{abi} L_{[b}\left(I_{j]L-1}^{(\ell+2)}+\left(\ell-1\right)I_{i_{\ell-1}]jL-2}^{(\ell+2)}\right)\biggr)^{\mathrm{TT}}.
\end{split}
\end{align}
Now, using the identity $\varepsilon_{abc}\varepsilon_{cde}=\delta_{ad}\delta_{be}-\delta_{ae}\delta_{db}$, which, in particular, implies $A_{[ab]}=(1/2)\varepsilon_{abc}\varepsilon_{cde}A_{de}$, one can rewrite (\ref{eq:identity3}) in the form
\be\label{eq:identity4}
\left(n_{aL-1}\varepsilon_{abi}I_{jL-1}^{(\ell+2)}L_{b}\right)^{\mathrm{TT}}=\left( n_{aL-1} \varepsilon_{abi}I_{(jL-1}^{(\ell+2)}L_{b)}+\frac{\ell-1}{\ell+1}n_{L-2}\varepsilon_{abi}I_{jaL-2}^{(\ell+2)}L_{b}\right)^{\mathrm{TT}},
\ee
where several terms proportional to $n_i$, $n_j$ and $\delta_{ij}$ vanish once the projection onto the TT gauge is applied. The first term on the rhs of (\ref{eq:identity4}) is much closer to the form we want. We just need to remove the traces of $I_{(jL-1}^{(\ell+2)}L_{b)}$. To this end, we use the identity
\be\label{eq:stf}
S_{L}=\hat{S}_{L}+\sum_{r=1}^{[\ell/2]}\frac{\left(-1\right)^{r+1}\ell!\left(2\ell-2r-1\right)!!}{\left(\ell-2r\right)!\left(2\ell-1\right)!!\left(2r\right)!!}\delta_{(i_{1}i_{2}}\cdots\delta_{i_{2r-1}i_{2r}}S_{i_{2r+1}\cdots i_{\ell})a_{1}a_{1}\cdots a_{r}a_{r}},
\ee
which expresses a fully symmetric tensor $S_L$ in terms of its STF part $\hat{S}_{L}\equiv S_{\langle L\rangle}$ plus traces \cite{Ross:2012fc}.
Note, however, that, using (\ref{eq:identity2}), one can show that tracing over any pair of indices of $I_{(jL-1}^{(\ell+2)}L_{b)}$ results in
\be\label{eq:trace}
\delta_{jb}I_{(jL-1}^{(\ell+2)}L_{b)}=\frac{2}{\ell+1}I_{aL-1}^{(\ell+2)}L_{a},
\ee
meaning that taking any further traces of (\ref{eq:trace}) yields zero, as $I_L$ is STF. Therefore, when applied to $I_{(jL-1}^{(\ell+2)}L_{b)}$, identity (\ref{eq:stf}) reduces to the single term $p=1$,
\be\label{eq:applied-stf}
I_{(jL-1}^{(\ell+2)}L_{b)}=I_{\langle jL-1}^{(\ell+2)}L_{b\rangle}+\frac{\ell}{2\ell+1}\delta_{(jb}I_{L-1)a}^{(\ell+2)}L_{a}.
\ee
Plugging (\ref{eq:applied-stf}) into (\ref{eq:identity4}), then (\ref{eq:identity4}) into (\ref{eq:e-stf}), using twice an identity similar to (\ref{eq:identity2}) and noting that many terms vanish due to either the presence of $\varepsilon_{abi}$, symmetrization over $ij$ or projection onto the TT gauge, one is able to arrive at
\begin{align}\label{eq:e-stf-2}
\begin{split}
\left[h_{ij}^{(e-L)}\right]^{\text{TT}}&=\frac{r^{-1}}{2}\biggl(A_{\ell}^{e}n_{aL-1}\varepsilon_{abi}I_{\langle jL-1}^{(\ell+2)}L_{b\rangle}+A_{\ell}^{e}\frac{\left(\ell-1\right)\left(\ell-2\right)}{\left(2\ell+1\right)\left(\ell+1\right)} n_{aL-3} \varepsilon_{abi} I_{jcbL-3}^{(\ell+2)}L_{c}\\
& +A_{\ell}^{e}\frac{\ell-1}{\ell+1}n_{L-2}\varepsilon_{abi}I_{jaL-2}^{(\ell+2)}L_{b} +B_{\ell}^{e}n_{L-2}L_{b}I_{aijL-3}^{\left(\ell+2\right)}\varepsilon_{i_{\ell-2}ab}+i\leftrightarrow j\biggr)^{\mathrm{TT}}.
\end{split}
\end{align}
The first term on the rhs of (\ref{eq:e-stf-2}) is now in the multipolar form that we want. Since $I_L$ is STF, the same is true of the second term, which contributes to $\mathcal{V}_{jbL-3}\propto I^{(\ell+2)}_{jcbL-3}L_c$. The remaining terms are worked out using exactly the same tools, which can be used to prove the following relations,
\begin{align}\label{eq:relations}
\begin{split}
&\pa{n_{L-2}\varepsilon_{abi}I_{jaL-2}^{\left(\ell+2\right)}L_{b}+i\leftrightarrow j}^{\mathrm{TT}}=\pa{n_{L-2}\varepsilon_{ab\langle i}I_{jL-2\rangle a}^{\left(\ell+2\right)}L_{b}
-\frac {\ell-2}{\ell}{ n_{a L-3}\varepsilon_{abi}I_{jcbL-3}^{\left(\ell+2\right)}L_{c} }+i \leftrightarrow j}^{\mathrm{TT}},\\
\\
&\pa{n_{L-2}L_{b}I_{aijL-3}^{\left(\ell+2\right)}\varepsilon_{i_{\ell-2} ab}+i\leftrightarrow j}^{\mathrm{TT}}=\pa{ n_{L-2}\varepsilon_{ab\langle i}I_{jL-2\rangle a}^{\left(\ell+2\right)}L_{b}
+\frac 2{\ell}{ n_{aL-3}\varepsilon_{a b i}I_{jcbL-3  }^{\left(\ell+2\right)}L_{c} }+i \leftrightarrow j}^{\mathrm{TT}}.\\
\end{split}
\end{align}
Plugging (\ref{eq:relations}) into (\ref{eq:e-stf-2}), one gets
\begin{align}
\begin{split}
\left[h_{ij}^{(e-L)}\right]^{\text{TT}} & =  \frac{r^{-1}}{2} \biggl\{ A_{\ell}^{e}n_{aL-1}\varepsilon_{abi}I_{\langle jL-1}^{(\ell+2)}L_{b\rangle}\\
& +\left[\frac{\left(\ell-1\right)\left(\ell-2\right)}{\ell+1}\left(\frac{1}{2\ell+1} - \frac {1}{\ell}\right)A_{\ell}^{e} +\frac 2{\ell} B_{\ell}^{e}\right] n_{aL-3}\varepsilon_{a b i}I_{jcbL-3  }^{\left(\ell+2\right)}L_{c}\\
& +\left(\frac{\ell-1}{\ell+1} A_{\ell}^{e}
+ B_{\ell}^{e}\right)n_{L-2}\varepsilon_{ab\langle i}I_{jL-2\rangle a}^{\left(\ell+2\right)}L_{b} +i\leftrightarrow j\biggr\}^{\mathrm{TT}},
\end{split}
\end{align}
where the last term contributes to $\mathcal{U}_{ijL-2}\propto \varepsilon_{ab\langle i}I_{jL-2\rangle a}^{\left(\ell+2\right)}L_{b}$.

\subsection{Current-type angular momentum tail}

For the angular momentum tail involving current-type multipole moments, we begin by writing (\ref{eq:e-m}) in a way similar to (\ref{eq:e-stf}),
\begin{align}\label{eq:m-stf}
\begin{split}
\paq{h^{(m-L)}_{ij}}^{\text{TT}} & = 16 r^{-1}\Bigl[A_{\ell}^m \pa{(1-\delta_{\ell2})n_{L-3}L_a J_{aijL-3}^{\pa{\ell+2}}-n_{aL-2}J_{ijL-2}^{\pa{\ell+2}}L_a}\\
& +B_{\ell}^m n_{aL-2}L_{(i} J_{j)aL-2}^{\pa{\ell+2}}\Bigr]^{\mathrm{TT}}.
\end{split}
\end{align}
Since $J_L$ is STF, the first term on the rhs of (\ref{eq:m-stf}) already has the structure of a mass-type radiative multipole moment $\mathcal{U}_{ijL-3}\propto J_{ijaL-3}^{(\ell+2)}L_a$. Meanwhile, the two remaining terms can be worked out using the tools of the mass-type case of Appendix~\ref{app:e-stf},
\begin{align}\label{eq:relations2}
\begin{split}
&n_{aL-2} J_{ijL-2}^{(\ell+2)}L_a=n_{L-1}L_{(i} J_{jL-1)}^{(\ell+2)}+\frac{2}{\ell+1}n_{aL-2}\varepsilon_{ab(i}J_{j)cL-2}^{(\ell+2)}\varepsilon_{bcd}L_{d},\\
&n_{aL-2}L_{(i} J_{j)aL-2}^{(\ell+2)}=n_{L-1}L_{(i}J_{jL-1)}^{(\ell+2)}- \frac{\ell-1}{\ell+1} n_{aL-2}\varepsilon_{ab(i}J_{j)cL-2}^{(\ell+2)}\varepsilon_{bcd}L_{d}.
\end{split}
\end{align}
In each of the cases of (\ref{eq:relations2}), we are left with a term containing $\varepsilon_{abi}$, which will eventually contribute to $\mathcal{V}_{jbL-2} \propto J_{c\langle jL-2}^{(\ell+2)}\varepsilon_{b\rangle cd}L_{d}$, as well as the term $n_{L-1}L_{(i} J_{jL-1)}^{(\ell+2)}$, which will eventually lead to a contribution to $\mathcal{U}_{ijL-1}\propto L_{\langle i} J_{jL-1 \rangle}^{(\ell+2)}$ once the traces are removed. Removing these traces with the aid of (\ref{eq:stf}) yields
\be
\pa{n_{L-1}L_{(i} J_{jL-1)}^{(\ell+2)}}^{\mathrm{TT}}=\pa{n_{L-1}L_{\langle i} J_{jL-1\rangle}^{(\ell+2)}+\frac{(\ell-1)(\ell-2)}{(2\ell+1)(\ell+1)}n_{L-3}L_{a} J_{aijL-3}^{(\ell+2)}}^{\mathrm{TT}}.
\ee
To extract the current-type radiative multipole moment from the other term (the one involving $\varepsilon_{abi}$), first we extract the fully symmetric part of $J_{jcL-2}^{(\ell+2)}\varepsilon_{bcd}L_{d}$ with the aid of the tools from Appendix~\ref{app:e-stf},
\begin{align}\label{eq:identity5}
\begin{split}
n_{aL-2}J_{jcL-2}^{(\ell+2)}\varepsilon_{bcd}L_{d} & =n_{aL-2}J_{c(jL-2}^{(\ell+2)}\varepsilon_{b)cd}L_{d}\\
& -\frac{1}{\ell}n_{aL-2}\left[\varepsilon_{bjc}J_{dcL-2}^{(\ell+2)}+(\ell-2)\varepsilon_{bi_{\ell-2}c}J_{dcjL-3}^{(\ell+2)}\right]L_{d}.
\end{split}
\end{align}
Once contracted with $\varepsilon_{abi}$ and projected onto the TT gauge, (\ref{eq:identity5}) becomes
\be
\left(n_{aL-2}\varepsilon_{abi}J_{jcL-2}^{(\ell+2)}\epsilon_{bcd}L_{d}\right)^{\mathrm{TT}}=\left(n_{aL-2}\varepsilon_{abi}J_{c\langle jL-2}^{(\ell+2)}\varepsilon_{b\rangle cd}L_{d}+\frac{\ell-2}{\ell}n_{L-3} L_{a} J_{aijL-3}^{(\ell+2)}\right)^{\mathrm{TT}},
\ee
where we note that the traces of $J_{c( jL-2}^{(\ell+2)}\varepsilon_{b) cd}L_{d}$ are automatically removed due to the presence of $\epsilon_{bcd}$ and the STF nature of $J_L$. Plugging (\ref{eq:relations2}) into (\ref{eq:m-stf}),
\begin{align}
\begin{split}
&\paq{h^{(m-L)}_{ij}}^{\text{TT}}= 16 r^{-1}\Bigl[ \left(B_{\ell}^m  -A_{\ell}^m \right) n_{L-1}L_{\langle i} J_{jL-1\rangle}^{(\ell+2)}\\
&+\left(\frac{(\ell-1)(\ell-2)}{(\ell+1)(2\ell+1)} \left(B_{\ell}^m  -A_{\ell}^m \right) +A_{\ell}^m-\frac{\ell-2}{\left(\ell+1\right)\ell}\left[2A_{\ell}^m  + \left(\ell-1\right) B_{\ell}^m\right]\right) \left(1-\delta_{\ell2}\right) n_{L-3} L_{a} J_{aijL-3}^{(\ell+2)} \\
&- \frac{2A_{\ell}^m  + \left(\ell-1\right) B_{\ell}^m}{\ell+1}\frac{1}{2}\left( n_{aL-2}\varepsilon_{abi}J_{c\langle jL-2}^{(\ell+2)}\varepsilon_{b\rangle cd}L_{d}+i\leftrightarrow j\right)\Bigr]^{\mathrm{TT}}.
\end{split}
\end{align}

\section{Family of integrals}
\label{app:ints}

The direct evaluation of the self-energy diagrams in Sec. \ref{se:results} can be done using the following family of two-loop integrals:
\begin{equation}
\label{eq:mi}
\int_{\K\Q}\frac1{{\cal D}_a}=\varepsilon \pa{-\omega^2}^{d-2-a}I_a\,,
\end{equation}
where the denominator ${\cal D}_a$ can be represented by ${\cal D}_a\equiv \pa{\K^2-\omega^2}\pa{\left(\K+\Q\right)^2-\omega^2}{\Q}^{2a}$ and the scalar $I_a$ is given by
\be
I_a\equiv\frac1{\pa{4 \pi}^d}\frac{\Gamma\pa{a+2-d}\Gamma\pa{a+1-d/2}^2\Gamma\pa{d/2-a}}{\Gamma\pa{2a+2-d}\Gamma\pa{d/2}}\,.
\ee
In Eq.~\eqref{eq:mi}, the coefficient $\varepsilon$ is either $\pm 1$, depending on the relative sign of the Feynman $i\varepsilon$-prescription in $\omega$ between in the propagators $(\K^2-\omega^2)$ and $\left(\left(\K+\Q\right)^2-\omega^2\right)$. For instance, for two retarded propagators, we have $(\K^2-(\omega+i 0^+)^2)$ and $(\left(\K+\Q\right)^2-(\omega+i 0^+)^2)$, and hence $\varepsilon=1$. Similarly, for one retarded and one advanced, like in $(\K^2-(\omega+i 0^+)^2)$ and $(\left(\K+\Q\right)^2-(\omega-i 0^+)^2)$, we have $\varepsilon = -1$.

\section{Unintegrated Amplitudes}
\label{app:unint}

In this  Appendix we collect the unintegrated expressions for the contributions to the effective action from the angular momentum tail self-energy.

We organize the expressions according to the field content in a Kaluza--Klein--type decomposition \cite{Kol:2007bc} of the metric perturbation, $h_{\mu\nu} \to \{\phi, A_i, \sigma_{ij}\}$. The labels in braces (e.g.\ $\{\sigma^2 A\}$) indicate schematically the number of fields of each type appearing in a given contribution.

We denote by $G_R(\mathbf{k},\omega)$ and $G_A(\mathbf{k},\omega)$ the retarded and advanced propagators, respectively, defined according to
\begin{equation}
G_{R}(\mathbf{k},\omega) \equiv \frac{1}{\mathbf{k}^2 - (\omega + i 0^+)^2}\qquad G_{A}(\mathbf{k},\omega) \equiv \frac{1}{\mathbf{k}^2 - (\omega - i 0^+)^2}.
\end{equation}

\subsection{Process $I^{ijR}\times J^{b\mid a}\times I^{klR^{\prime}}$}

\begin{align}
i S_{\rm eff} &= \frac{(-1)^{r+1} i^{r+r'}c_r^{(I)} c_{r'}^{(I)}}{16\Lambda^4} \int \frac{d\omega}{2\pi} \int_{\K\Q} \frac{(k+q)_R k_{R'}q_a}{\Q^2} \nonumber\\
&\times [
  J^{b \mid a}_+ I^{ijR}_+(\omega) I^{klR'}_-(-\omega) G_R(\K+\Q,\omega) G_R(\K,\omega)
+ J^{b \mid a}_+ I^{ijR}_-(\omega) I^{klR'}_+(-\omega) G_A(\K+\Q,\omega) G_A(\K,\omega) \nonumber\\
&+ J^{b \mid a}_- I^{ijR}_+(\omega) I^{klR'}_+(-\omega) G_R(\K+\Q,\omega) G_A(\K,\omega)] \nonumber\\
&\times \Bigg\{ \frac12 \omega^5 [\delta_{jl} (2k_b \delta_{ik} + q_b \delta_{ik} +2q_i \delta_{kb}) -2q_l \delta_{ik}\delta_{jb}] \qquad \{ \sigma^2 A\} \nonumber\\
&\quad+\omega^3 [k_k k_l q_i \delta_{jb}-(k+q)_i (k+q)_j q_k \delta_{lb}] \qquad \{ \sigma A \phi\} \nonumber\\
&\quad-\omega^3 (k+q)_j k_l [(2k_b+q_b)\delta_{ik} - q_i \delta_{kb} + q_k \delta_{ib}] \qquad \{ A^3\} \nonumber\\
&\quad+ \omega^3 \Big\{ (k+q)_j [ (k_i q_k - k_k q_i)\delta_{bl} + (k_l q_b- k_b q_l)\delta_{ik} + (k+q)_k q_l \delta_{ib} + (\K+\Q)\cdot\Q \delta_{bl} \delta_{ik} ] \nonumber\\
&\quad\qquad-k_l [ (k_k q_j - k_j q_k) \delta_{ib} + (k_j q_b - k_b q_j)\delta_{ik} +k_i q_j \delta_{kb}
+\K\cdot\Q \delta_{ik} \delta_{jb} ]
\Big\} \qquad \{ \sigma A^2\} \nonumber\\
&\quad+\omega (k+q)_j k_l [ (k+q)_i k_b q_k - k_k q_b q_i - k_b k_k q_i -(k+q)_i \K\cdot\Q \delta_{kb} +k_k (\K+\Q)\cdot\Q \delta_{ib} ]  \qquad \{ A^2 \phi \} \nonumber\\
&\quad+\frac{\omega}{c_d} (k+q)_i (k+q)_j k_k k_l (2k_b + q_b)  \qquad \{ A \phi^2 \}
\Bigg\}\,.
\end{align}

\subsection{Process $J^{b\mid iRa}\times J^{m\mid n}\times J^{d\mid jR^{\prime}c}$}

\begin{align}
i S_{\rm eff} &= \frac{(-1)^{r+1} i^{r+r'+2}c_r^{(J)} c_{r'}^{(J)}}{64\Lambda^4} \int \frac{d\omega}{2\pi} \int_{\K\Q} \frac{(k+q)_R k_{R'} k_c q_n}{ \Q^2} \nonumber\\
&\qquad\qquad\times [
  J^{m \mid n}_+ J^{b \mid iRa}_+(\omega) J^{d \mid jR'c}_-(-\omega) G_R(\K+\Q,\omega) G_R(\K,\omega) \nonumber\\
&\qquad\qquad+ J^{m \mid n}_+ J^{b \mid iRa}_-(\omega) J^{d \mid jR'c}_+(-\omega) G_A(\K+\Q,\omega) G_A(\K,\omega) \nonumber\\
&\qquad\qquad+ J^{m \mid n}_- J^{b \mid iRa}_+(\omega) J^{d \mid jR'c}_+(-\omega) G_R(\K+\Q,\omega) G_A(\K,\omega)] \nonumber\\
&\times \Bigg\{ \omega^3 (k+q)_b 
\big[ 2k_m (\delta_{ad}\delta_{ij} + \delta_{id}\delta_{ja}) + q_i (\delta_{ad}\delta_{jm} + \delta_{dm}\delta_{ja}) - q_j (\delta_{am}\delta_{id} + \delta_{ad}\delta_{im}) \nonumber\\
&\qquad\qquad\qquad\qquad\quad+ q_m (\delta_{ad}\delta_{ij} + \delta_{id}\delta_{ja}) + q_a (\delta_{id}\delta_{jm} + \delta_{dm}\delta_{ij}) - q_d (\delta_{am}\delta_{ij} + \delta_{im}\delta_{ja}) \big]\qquad \{ \sigma^2 A  \} \nonumber\\
&\qquad\,\, +\omega (k+q)_b \Big\{ k_j \big[ (k_m q_i - k_i q_m) \delta_{ad} +(k_m q_a - k_a q_m) \delta_{id} - (k_i q_a + k_a q_i) \delta_{dm} \nonumber\\
&\qquad\qquad\qquad\qquad\qquad\,\,\,+ k_i q_d \delta_{am} + k_a q_d \delta_{im} 
-\K\cdot\Q (\delta_{am}\delta_{id} + \delta_{ad}\delta_{im}) \big] \nonumber\\
&\qquad\qquad\qquad+(k+q)_i \big[ (k_j q_m - k_m q_j) \delta_{ad} +(k_j q_d + 2 q_j q_d) \delta_{am} - k_m q_d \delta_{ja} + k_a q_d \delta_{jm} \nonumber\\
&\qquad\qquad\qquad\qquad\qquad\,\,\,+ (k_a q_j - k_j q_a) \delta_{dm}  
+(\K+\Q)\cdot\Q (\delta_{dm}\delta_{ja} + \delta_{ad}\delta_{jm}) \big] 
 \Big\} \qquad \{ \sigma A^2  \} \nonumber\\
&\qquad\,\, +\omega (k+q)_{i} (k+q)_{a} k_j  \big( 2 k_m\delta_{bd} + q_m \delta_{bd} + q_d \delta_{bm} - q_b\delta_{dm} \big) \qquad \{  A^3  \} \Bigg\}
\end{align}

\subsection{Process $I^{ijR}\times J^{m\mid n}\times J^{d\mid jR^{\prime}c}$}

\begin{align}
i S_{\rm eff} &= \frac{(-1)^{r+1} i^{r+r'+2}c_r^{(I)} c_{r'}^{(J)}}{16\Lambda^4} \int \frac{d\omega}{2\pi} \int_{\K\Q} \frac{(k+q)_R k_{R'} k_a q_n}{ \Q^2} \nonumber\\
&\qquad\qquad\times [
  J^{m \mid n}_+ I^{ijR}_+(\omega) J^{b \mid kR'a}_-(-\omega) G_R(\K+\Q,\omega) G_R(\K,\omega) \nonumber\\
&\qquad\qquad+ J^{m \mid n}_+ I^{ijR}_-(\omega) J^{b \mid kR'a}_+(-\omega) G_A(\K+\Q,\omega) G_A(\K,\omega) \nonumber\\
&\qquad\qquad+ J^{m \mid n}_- I^{ijR}_+(\omega) J^{b \mid kR'a}_+(-\omega) G_R(\K+\Q,\omega) G_A(\K,\omega)] \nonumber\\
&\times \Bigg\{ 
\omega^4 \big[ (2k_m+q_m)\delta_{ik} \delta_{jb} + q_j (\delta_{ib}\delta_{km}+\delta_{bm}\delta_{ik}) - (q_k \delta_{jb} + q_b \delta_{jk}) \delta_{im} \big] \qquad \{ \sigma^2 A \}
\nonumber\\
&\quad + \omega^2 k_k (k+q)_j (q_i \delta_{bm} - q_b \delta_{im} - 2k_m \delta_{ib}-q_m \delta_{ib}) \qquad \{ A^3 \} \nonumber\\
&\quad - \omega^2 (k+q)_i (k+q)_j (q_k \delta_{bm} + q_b \delta_{km}) \qquad \{ \sigma A \phi \} \nonumber\\
&\quad -  k_k (k+q)_i (k+q)_j (k_b q_m - k_m q_b + \K\cdot\Q \delta_{bm}) \quad \{ A^2 \phi \} \nonumber\\
&\quad + \omega^2  \big[ ( k_i k_j q_k + k_i q_j q_k - 2 k_i k_k q_j - k_k q_i q_j) \delta_{bm} + (k_k k_m q_j + k_k q_j q_m - (k+q)_j k_m q_k) \delta_{ib} \nonumber\\
&\qquad\quad- k_m q_b (k+q)_j \delta_{ik} + (2k_j k_k + 2k_j q_k + k_k q_j + 2q_j q_k) q_b \delta_{im} + k_i q_b (k+q)_j \delta_{km} \nonumber\\
&\qquad\quad+(k+q)_j (\K+\Q)\cdot\Q (\delta_{bm} \delta_{ik} + \delta_{ib} \delta_{km}) - k_k \K\cdot\Q \delta_{ib} \delta_{jm} \big] \qquad \{ \sigma A^2 \}
\Bigg\}
\end{align}

\end{document}